\documentclass[prc,onecolumn,tightenlines,12pt]{revtex4}
\usepackage{graphicx}
\usepackage{dcolumn}
\usepackage{amsmath}
\newcommand{\be}{\begin{equation}}
\newcommand{\ee}{\end{equation}}
\newcommand{\bea}{\begin{eqnarray}}
\newcommand{\eea}{\end{eqnarray}}

\begin{document}

\author{M. S. Hussein$^{\dag 1,2}$ }
\affiliation{$^{1}$Max-Planck-Institut f\"ur Physik komplexer Systeme\\
N\"othnitzer Stra$\beta$e 38, D-01187 Dresden, Germany \\
$^{2}$Instituto de F\'{i}sica, Universidade de S\~{a}o Paulo\\
C.P. 66318, 05315-970 S\~{a}o Paulo, S.P., Brazil}
\title{Absorption-Fluctuation Theorem for Nuclear Reactions: Brink-Axel, Incomplete Fusion and All That
\thanks{Supported in part by the CNPq and FAPESP (Brazil).\\
        $^{\dag}$Martin Gutzwiller Fellow, 2007/2008.}}

\begin{abstract}
We discuss the connection between absorption, averages and fluctuations in nuclear reactions.
The fluctuations in the entrance channel result in the compound nucleus, Hauser-Feshbach, cross section,
the fluctuations in the intermediate channels, result in modifications of multistep reaction cross
sections, while the fluctuations in the final channel result in hybrid cross sections that can be
used to describe incomplete fusion reactions. We discuss the latter in details and comment on the
validity of the assumptions used in the develpoment of the Surrogate method. We also discuss the
theory of multistep reactions with regards to intermediate state fluctuations and the energy dependence 
and non-locality of the intermediate channels optical potentials.
\end{abstract}
\maketitle

\newpage

\section{Introduction}

In the theory of nuclear reactions, one relies on statistical ideas to simplify their many-body
nature. The first step is to introduce average amplitudes. The second step is to devise a Schr\"odinger 
equation that supplies these amplitues (the optical model). and the last step is the calculation
of the fluctuation cross section. Several methods are available that supplies the derivation of this 
fluctuation cross section. The concept of energy averages and/or ensemble averages are used to obtain
the Hauser-Feshbach cross section and corrections to it. The ergodic theorem is commonly employed to
argue the equivalence of the two averaging methods. As much as fluctuations arise when one uses average
amplitudes, the concept of fluctuations can be generalized to multistep processes, where the intermediate channel
green functions are replaced by their averages. Similarly, if an exit doorway dominates the reaction,
such as the excitation of a giant resonance in one of the nuclei participating in the reaction, or the 
coupling to the continuum, conveniently discretized, the final channel fluctuations set in resulting
in an incomplete fusion cross section  In both of these cases, the absorption and fluctuation analysis is
made on the exact entrance channel and exit channel wave functions.

In multistep reactions, another important reaction operator comes into the picture. The many-body 
intermediate channel Green's function. Again, one can decompose this quantity into an optical piece
and a fluctuation piece. The fluctuation part can be calculated using the different optical quantitites
as they appear in the KKM Optical Background Representation. One important addition to this procedure
is the inclusion of other collective states which might be excitated in the intermediate propagation 
of the system. We have in mind the Brink-Axel mechanism. This has been accomplished recently in the
theoretical description of the excitation of multiphonon states. It is important to extend this new feature
to multistep direct reactions in general. 

In this talk I will discuss the incomplete fusion reaction theory and its connection to the Surrogate Method. I also discuss the Brink-Axel effect in multistep reaction theory. In both cases the Absorption- Fluctuation "theorem" is invoked to get a practical closed theory of the reactions. Some comments concerning the foundation of the Statistical Multistep
Reaction Theory are also made.

The wave function fluctuation can be written as

$\Psi=\Psi_{optical} + \Psi_{fluctuation}$

where $\Psi_{fluctuation}$ can be related to the optical quantities contained in $\Psi=\Psi_{optical}$. By construction
the energy average of the latter wave function is $0$. This then results in an average cross section containing
an optical one ( calculated with DWBA or Coupled Channels Theory) and a fluctuation cross section calculated with
optical transmission coefficients ( or matrices). If the wave function refers to the final channel, then the resulting
cross section will contain a direct one plus a fluctuation one which refers to the formation of a compound nucleus in a subsystem ( if one has in mind a breakup process a(b +x) + A $\rightarrow$ b + (x +A) $\rightarrow$ b + c + C. The total formation cross section of the compound subsystem, a(b +x) + A $\rightarrow$ b + (x +A), is the incomplete fusion cross section.

When dealing with multistep processes, the intermediate channel Green's function, $G_{i}$ can be decomposed into an optical
plus a fluctuation pieces. The optical part contains complex energies. The fluctuation one is a multistep Green's function itself. It is this latter which would, if the conditions are met, contain the Brink-Axel effect.

\section{ Final State Fluctuations}

We consider the following reaction \cite{HM85, HMa89, HFM90, Fr91}, \cite{KM79, Par95}, \cite{Ud84, Au87}

\begin{equation}
a(x+b) +A \rightarrow b + \sum_{all states}(x + A),
\end{equation}
and treat b as a spectator, namely it only suffers elastic scattering in the optical potential field supplied by the target.

The cross-section for observing b is ( no CN formation in the xA subsystem ),

\begin{equation}
\frac{d\sigma_{b}^{(direct)}}{d\Omega_{b}dE_{b}}=\frac{2}{\hbar v_{a}} \rho(E_{b}) < \rho_{x}^{(+)}(\mathbf{r}_{x})|W_{xA}^{(D)}(\mathbf{r}_{x}, E_{i} +B_{a}-E_{b})| \rho_{x}^{(+)}(\mathbf{r}_{x}) >
\end{equation}
where $\rho(E_{b})$ is the density of final b-states,
$\rho(E_{b}) = \frac{\mu_{b}k_{b}}{(2\pi)^{3}\hbar^{3}}$,
and the "negative-energy entrance channel wave function" of x, $\rho_{x}^{(+)}(\mathbf{r}_{x})$ is,
\begin{equation}
\rho_{x}^{(+)}(\mathbf{r}_{x})=(\psi_{b}^{(-)}(\mathbf{r}_{b}) | \phi_{a} (\mathbf{r}_{b}- \mathbf{r}_{x}) \Psi_{a}^{(+)}(\mathbf{r}_{b},\mathbf{r}_{x}))
\end{equation}

where $\Psi_{a}^{(+)}$ is the full three-body wave function of the $b+x+A$ system. Further, the imaginary potential $W_{xA}^{(D)}(\mathbf{r}_{x}, E_{i} +B_{a}-E_{b})$ takes into account the direct reactions of x with the target.

The above equation for the direct inclusive cross section cross does not take into account the fluctuation in the xA subsystem owing to the CN formation. The inclusion of this contribution can be easily made by adding another piece to the imaginary part of the xA optical potential. This is easily seen if we write the cross section for observing b as

\begin{eqnarray}
\frac{d\sigma_{b}}{d\Omega_{b}dE_{b}} = \frac{2\pi}{\hbar v_{a}}\rho(E_{b})\sum_{c} 
<\Psi^{(+)}(\mathbf{r_{x}},\mathbf{r_{b}})\Phi_{A}| V_{bx}|\psi_{b}^{(-)}\Psi_{xA}^{c}> 
\delta(E-E_{b}-E^{c})\times & \nonumber\\
<\psi_{b}^{(-)}\Psi_{xA}^{c}| V_{bx}|\Psi^{(+)}(\mathbf{r_{x}},\mathbf{r_{b}})\Phi_{A}>.
\end{eqnarray}

The above equation can be reduced using standard procedures into the following compact form

\begin{equation}
\frac{d\sigma_{b}}{d\Omega_{b}dE_{b}} = \frac{d\sigma_{b}^{elastic-breakup}}{d\Omega_{b}dE_{b}} + \frac{d\sigma_{b}^{inelastic-breakup}}{d\Omega_{b}dE_{b}}
\end{equation}

where the inelastic breakup cross section can be further decomposed into,

\begin{equation}
\frac{d\sigma_{b}^{inelastic-breakup}}{d\Omega_{b}dE_{b}} = \frac{d\sigma_{b}^{(direct)}}{d\Omega_{b}dE_{b}} + \frac{d\sigma_{b}^{(incomplete-fusion)}}{d\Omega_{b}dE_{b}}.
\end{equation}

From the general form of the imaginary potential one can write the following for the incomplete fusion cross section which is the final state fluctuation contribution

\begin{equation}
\frac{d\sigma_{b}^{(incomplete-fusion)}}{d\Omega_{b}dE_{b}} = \rho(E_{b})\sum_{c} \frac{v_{c}}{v_{a}}\sigma_{F}^{ac}( E_{i} +B_{a} -E_{b})
\end{equation}

where $\sigma_{F}^{ac}$ is the fusion cross section from channel c ( corresponding to an excited non-elastic channel of the xA subsystem), which is populated by a. 

\begin{equation}
\sigma_{F}^{ac} = \frac{k_{c}}{E_{c}}<\rho_{c}^{(+)}|W_{c}|\rho_{c}^{(+)}>.
\end{equation}
with
\begin{equation}
|\rho_{c}^{(+)}> = G_{p_{c}}^{(+)opt} V_{p_{c}P}|\rho_{x}^{(+)}>.
\end{equation}

Cleraly, Eq.(6) can be written more generally as, after introducing the $xA$ fusion imaginary potential, $W_{xA}^{(F)}$,
\begin{equation}
\frac{d\sigma_{b}^{inelastic-breakup}}{d\Omega_{b}dE_{b}}=\frac{2}{\hbar v_{a}} \rho(E_{b}) < \rho_{x}^{(+)}(\mathbf{r}_{x})|[W_{xA}^{(D)}(\mathbf{r}) + W_{xA}^{(F)}(\mathbf{r})]| \rho_{x}^{(+)}(\mathbf{r}_{x}) >
\end{equation}

Eq. (10), with $W^{(F)}$ associated with the annihilation component, was used to calculate the inclusive annihilation of very low energy antiprotons on deuterium \cite{Fr90}. Further, the above theory has been also used to treat incomplete fusion reactions involving weakly bound exotic nuclei \cite{Can98}.
Of course direct capture of x, without the concomitant excitation of the subsystem
is the first term in the sum above. It is the above incomplete fusion cross section which is invariably referred to as hybrid, Trojan-Horse \cite{TB03} or Surrogate \cite{Ple05}, \cite{ED06}, cross section. In all of these applications only the direct capture is taken into account. Further, the x-particle is taken to be a geniune projectile, without due attention to the fact that its energy is dispersed by the internal, Fermi energy, inside a, the carrier of x.

\section{General Characteristics of Hybrid Reactions}

In the Surrogate method one uses a description of a "Desired reaction", $\mathbf{Dr}$ in terms of a "Surrogate reaction", $\mathbf{Sr}$. The $\mathbf{Sr}$ is invariably of the type
$d + D \rightarrow b + B^{*} \rightarrow b + c + C + ....$ and the $\mathbf{Dr}$ of the type $a + A \rightarrow B^{*} \rightarrow c + C + ....$\cite{Esc07}. The compound reaction in the $\mathbf{Dr}$ is described by the Hauser-Feshbach theory, which supplies the following for the cross section, 

\begin{equation}
\sigma_{\alpha\chi} = \sum_{J,\pi} \sigma_{\alpha}^{CN}(E,J,\pi) \cdot G^{CN}_{\chi}(E,J,\pi),
\end{equation}

where $\sigma_{\alpha}^{CN}$ is the partial fusion cross section ( CN formation cross section) and can be calculated from the optical model, while
the function $G^{CN}_{\chi}$ is the branching ration for the CN decay. This latter quantity is difficult to predict. For this, one relies on the $\mathbf{Sr}$,
which gives,

\begin{equation}
P_{\chi}(E) = \sum_{J,\pi} F_{\delta}^{CN}(E,J,\pi) \cdot G^{CN}_{\chi}(E,J,\pi),
\end{equation}

where the function $F_{\delta}^{CN}(E,J,\pi)$ is evaluated from direct reaction theory and is intimately connected to the partial incomplete fusion cross section of Eqs.(7-9). What is usually done, however, is to assume $(J,\pi)$-independent $G^{CN}$ and employ the Weisskopf-Ewing description of the $\mathbf{Dr}$ and the $\mathbf{Sr}$ ($P_{\chi}(E) = G^{CN}_{\chi}(E)$). This prescription finally gives,

\begin{equation}
\sigma_{\alpha,\chi}^{WE}(E) = \sigma_{\alpha}^{CN}(E) \cdot P_{\chi}(E).
\end{equation}

To date, most of the effort in the application of the $\mathbf{Sr}$ relied on the WE approximation and Eq. (13). The Surrogate Method has been quite
successful for a variety of $\mathbf{Sr}$ and it would be only natural to asky why, since several rather drastic assumptions (e.g. the WE approximation, and the use of on-shell description) are employed. In the following, I raise several theoretical issues.

1- "The "incident projectile" energy, $E_{a}$ is dispersed by an amount directly related to its internal, Fermi, motion
inside the real projectile $d$. The wave function of the captured particle $a$ is given by the source wave function, $\rho_{a}^{(+)}(\mathbf{r}_{a})$, which describes the incoming $a$ off the energy shell. If the dispersion in the $a$ energy is ignored as done in the Surrogate Method, one may wonder how large the off shell effects would be.

2- The capture of $a$ into the $a-A$ compound nucleus $B^{*}$ proceeds in a multistep fashion. This is quite apparent from the general structure of the fusion cross sections appearing inside the sum above.

3- The critical questions in Surrogate Methods are not the CN decay ratios. Rather, it is the formation of the CN.

4- The d particle could be a " phonon-photon" as in $(\alpha,\alpha^{\prime})$ reactions. How to treat these in 
view of the theory above? One may envisage using the equivalent nuclear-phonon method, as develpoed by Feshbach and Zabek
\cite{FZ77}

\section{Intermediate State Fluctuations}

In the excitation of nuclear states through multistep processes one may find it necessary to take into account the fluctuations inherent in the intermediate propagation of the system. This can be important in the excitation of multiphonon states \cite{Car99, CH99, Can99, Hu02}, \cite{GW01}. If the collision time is comparable to the decay time of the intermediate collective state, then through the damping one may excite a collective Brink-Axel phonon on top of the background of the one phonon state. This will add
a different type of fluctuation cross section, which has to be added to the usual mutistep cross section. As a result,
the low-energy multistep direct cross section has to be modified. The basic quantity that averages and consequently
develops fluctuations is the intermediate Green function. To see how the intermediate state fluctuations affects the cross section we first give a very brief description of the \cite{FKK80} statistical multistep direct reaction theory.

The n-step transition from the initial channel  $i$ to the final channel $f$ through the action of a transition potential operator $V$ is given by an amplitude which has the general form,

\begin{equation}
T_{if} = V_{fn-1} G^{(+)}_{n-1}(E)\cdots V_{32} G^{(+)}_{2}(E) V_{21}G^{(+)}_{1}(E)V_{1i}
\end{equation}

where the intermediate channel Green's function $G^{(+)}_{j}(E)$ is given by,

\begin{equation}
G^{(+)}_{j}(E)= \frac{1}{ E - H_{j} + i\epsilon}
\end{equation} 

where $H_{j}$ is the many-channel (body) Hamiltonian describing the colliding nuclear system in channel $j$.

To be specific, we  consider the excitation of multiphonon giant resonances. In this case the channels $j$ refer to doorway channels,
with $d_{1}$ being the the first doorway or GDR, while $d_{2}$ the second doorway or the Double Giant Dipole resonce (DGDR)
etc. The propagator in the space of $d_{1}$ is then denoted by the Green's function $G^{(+)}_{d_{1}}(E)$.

This Green function can be written as:

\begin{equation}
G^{(+)}_{d_{1}}(E) = \overline{G^{(+)}_{d_{1}}}(E) + G^{(+) fluct.}_{d_{1}}(E)
\end{equation}

where $\overline{G^{(+)}_{d_{1}}}(E)$ is the average Green function containing the damping width of the intermediate one-phonon collective state viz,

\begin{equation}
\overline{G^{(+)}_{d_{1}}}(E) = \frac{1}{E-H_{1}-\varepsilon_{1}+i\frac{\Gamma_{d_{1}}^{\downarrow}}{2}}
\end{equation}

and the fluctuation contribution contains explicit reference to the fine structure states that give rise to $\Gamma_{{d}_{1}}^{\downarrow}$
\begin{equation}
G^{(+) fluct.}_{d_{1}}(E) = G_{q_{1}}(E)v_{q_{1}d_{1}}\overline{G^{(+)}_{d_{1}}}(E)
\end{equation}

The Green function $G_{q_{1}}$ represents propogation of the system in the $q_{1}$ subspace, the latter being spanned by the fine structure states to which the first doorway ( the single-phonon resonance) is damped and as a consequence it acquires the damping width. When calculating the two-step cross section, $|0> \rightarrow |d_{1}> \rightarrow |d_{2}>$,
one would obtain the average one, related to $\overline{G^{(+)}_{d_{1}}}(E)$ plus a fluctuation one related to
$G^{(+) fluct.}_{d_{1}}(E)$. This latter can be evaluated using the usual energy averaging procedure of products of two
rapidly fluctuating Green functions; $G_{q_{1}}(E) G_{q_{1}^{\prime}}^{\dagger}(E^{\prime})$. The resulting cross section
is proportional to $\Gamma_{d_{1}}^{\downarrow} \tau_{c}(E)$, where $\tau_{c}(E)$ is the collision time. If we call the average cross section of the two-step process, $\overline\sigma^{(2)}$, the usual multistep (MS) cross section for going from the ground state to the one-phonon doorway followed by the transition to the final, two-phonon doorway, through $\overline{G^{(+)}_{d_{1}}}(E)$, by
$\sigma_{MS}^{(2)}$, and the two-step fluctuation cross section, by $\sigma_{fl}^{(2)}$, then we find,

\begin{equation}
\sigma_{fl}^{(2)} = \frac{\Gamma_{d_{1}}^{\downarrow}\tau_{c}(E)}{2\hbar}\cdot{\sigma_{MS}^{(2)}}
\end{equation}

Accordingly the energy averaged cross section is

\begin{equation}
\overline\sigma^{(2)}= \left( 1 + \frac{\Gamma_{d_{1}}^{\downarrow}\tau_{c}(E)}{2\hbar}\right)\cdot{\sigma_{MS}^{(2)}}
\end{equation}

It is evident that the fluctuation contribution to the two step cross section resulting from the excitation of the Brink-Axel phonon on top of the background of the one-phonon doorway, could become important at energies where the collision time is appreciable ( low energies). In the case of the excitation of the double giant dipole resonance (DGDR)
in the reaction $^{208}$Pb + $^{208}$Pb at 640 MeV.A bombarding energy the fluctuation contributio is found to
contribute 30$\%$ to the cross section \cite{CH99}. At the lower energy of 100 MeV.A, the contribution is 100$\%$. The same findings were reported in \cite{GW01} where the method of supersymmetry ensemble average \cite{VZW85} was employed to obtain the Brink-Axel fluctuation term in the two-step cross scetion. The method of \cite{Can99} can also be used to derive the cross section for the three-step (TGDR) process \cite{Hu02} and the general form of the n-step cross section, viz
\begin{equation}
\overline\sigma^{(n)}= \big[ 1 + \sum_{k=1}^n \frac{(n-k)!}{n!}\frac{(n-k)}{(n+k)}\binom{n}{k}\left(\frac{\Gamma_{d_{1}}^{\downarrow}\tau_{c}(E)}{\hbar}\right)^{k}\big]
\cdot{\sigma_{MS}^{(n)}}
\end{equation}
It is therefore clear that intermediate state fluctuations must be given due considerations in the theory of multistep
direct reactions. It remains to be seen what has to be done in the SMSD theory of \cite{FKK80} in order to take into account the intermediate state fluctuations and assess the convergence of the resulting modified expression of the cross section.

\section{Non-Locality, Energy-Dependence and Complexity of the Optical Potential} 

The optical potential is the backbone of nuclear reaction theory. The original motivation behind the introduction of the
OP was the need to analyse neutron scattering data with the one-body Schr\"odinger equation. The general properties
of the OP can be summarized by three basic characteristics: Energy-dependence, Non-locality and Complexity. All of these
features are clearly exhibited in the Feshbach form of the Optical Potential Operator,
\begin{equation}
PU(E)P = PVQ G_{QQ}(E)QVP  
\end{equation}

where V couples the P-space to the Q-space and $G_{QQ}(E)$ is the full Green function in the Q-space. Here the P-space and the Q-space could be the open channels and closed channels (CN) subspaces, respectively, or simply two sets of open channels. In the case that Q represents closed channels, one resorts to energy averaging to smooth out the fluctuations in $G_{QQ}(E)$ due to the CN resonances. This can be expediently accomplished with the Kerman-Kawai-McVoy Optical Background Representation \cite{KKM73}. The energy averaged $G_{QQ}(E)$ is then just
\begin{equation}
\overline{G_{QQ}}(E) = G_{QQ}(E + iI)
\end{equation}
where I is the averaging energy interval. It is then evident that the complexity of $PU(E)P$ arises from the average intermediate state propagation in the Q-subspace. The optical potential is non-local owing to the same intermediate
state propagation,
\begin{equation}
PU(E,\mathbf{r},\mathbf{r}^{\prime})P = PV(\mathbf{r})QG_{QQ}(E + iI)QV(\mathbf{r}^{\prime})P
\end{equation}
If the Q-subspace represents other open channels not contained in P, the same arguments as above apply, with the Q-channel
propogator now acquiring an imaginary part owing to the boundary condition of open channels ( the $i\epsilon$ factor).

Evidently, non-locality, complexity and energy-dependence of the Feshbach optical potential originate from the same
source; the intermediate, Q-space, propagation of the system. Of course, the non-locality arising from the exchange effects
should also be present, through the addition to the Feshbach potential, what is commonly called the bare potential, which is
taken to be real. This is the mean field potential acting at positive energies. In actual applications in nuclear reactions
of relevance to this Workshop, the optical potential is usually taken to be local, complex and with slow energy dependence.
Obtaining an equivalent local potential from the non-local one is justified using the Perey-Buck\cite{PB62} prescription, which brings in a non-despersive energy dependence besides the dispersive one contained in the intermediate Q-space propagator ( see above ). 

The presence of energy dependence in the intermediate channel optical potential
which appears in the Green function brings in problems related to the contruction of the dual states in the spectral decomposition of the latter. These dual scattering states can not be obtained from the optical Schr\"odinger equation by merely changing the optical potential by its complex adjoint as is customary \cite{HB82}, \cite{KA91}, \cite{Ar95}. In fact, the dual scattering states for energy-dependent optical potentials can only be formally obtained by solving an integral equation
containing in its kernel the half-on-energy-shell physical $T$-matrix. Of course when formulating
the MSD reaction without the use of the spectral decomposition of the intermediate channel Green function as was done in
\cite{TUL83}, \cite{NWY88} for two-step processes, the above question of the dual state is avoided. On the other hand, to extend such theories to processes involving more steps is rather impractical and the use of procedures such as those of \cite{FKK80} become unavoidable as long as the correct handling of the dual states is employed. First we recall the approximations employed by \cite{FKK80} to obtain, from Eq. (14), a cross section for multistep processes which is given
by a convolution of single step, DWBA-like, cross sections. Let us suppose that the intermediate channel $j$ Green's function is replaced by its energy averaged part, and call it $\overline{G^{(+)}_{j}}(E)$. This Green's function is given by

\begin{equation}
\overline{G^{(+)}_{j}}(E)= \frac{1}{ E - H_{0} - U_{j}(E) + i\epsilon}
\end{equation} 

In the above, the optical potential $U_{j}(E)$ is, as already alluded above, energy-dependent and consequently complex, non-local potential. The three basic assumptions of \cite{FKK80} are: a) ingnore the energy-dependence of the optical potential but keep it complex, b) use the spectral representation of the resulting Green's function in the form,

\begin{equation}
\overline{G^{(+)}_{j}}(E) = \int\frac{d\mathbf{k}}{(2\pi)^{3}}\frac{|\Psi_{\mathbf{k}}^{(+)}> <\widetilde{\Psi}_{\mathbf{k}}^{(+)}|}{ E - \varepsilon_{j} - E_{\mathbf{k}}+ i \epsilon}
\end{equation}

and, c) take only the on-energy-shell part of the above ( the delta function part), viz,

\begin{equation}
\overline{G^{(+)}_{j}}(E) \approx -i\pi \int \frac{d\mathbf{k}}{(2\pi)^{3}} \delta( E_{j} - E_{\mathbf{k}}) |\Psi_{\mathbf{k}}^{(+)}> <\widetilde{\Psi}_{\mathbf{k}}^{(+)}| = \frac{i}{2\pi}(\frac{2\mu}{\hbar^{2}})^{3/2}E_{j}^{1/2}
|\Psi_{\mathbf{k_{j}}}^{(+)}> <\widetilde{\Psi}_{\mathbf{k_{j}}}^{(+)}|
\end{equation}

where $E_{j} = E -\varepsilon_{j}= \frac{\hbar^{2}k_{j}^{2}}{2\mu}$

With the above form of the Green's function, the multistep series can be collapsed into a product of one-step amplitudes of the type,

\begin{equation}
\widetilde{T}_{12} = <\widetilde{\Psi}_{\mathbf{k_{2}}}^{(+)}| V_{12} |\Psi_{\mathbf{k_{1}}}^{(+)}>
\end{equation}

The above amplitude is not of the usual DWBA type as the dual solution $<\widetilde{\Psi}_{\mathbf{k_{2}}}^{(+)}|$ appears
instead of the usual $<\Psi_{\mathbf{k_{2}}}^{(-)}|$. Only the last term in the series involving the transition to the final channel is of the usual DWBA form. In the original FKK paper \cite{FKK80} such a distinction was not made and all terms in the series were taken to be of the DWBA type. This assumption, however,  was corrected later by \cite{KA91}.

In the following we argue that the energy dependence of the intermediate channel Green's function, even if it is weak, does not allow one to use the dual wave function above, obtained in \cite{KA91} and \cite{Ar95} by solving the optical Schr\"odinger equation with $U_{j}$ replaced by its complex adjoint $U^{\dagger}_{j}$.

Consider the Schr\"odinger equation with energy-dependent interaction

\begin{equation}
(E_{k} - H_{0} - U(E_{k})) |\Psi_{\mathbf{k}}^{(+)}> = 0
\end{equation}

The dual scattering state is then defined such that

\begin{equation}
<\widetilde{\Psi}_{\mathbf{k}}^{(+)}|\Psi_{\mathbf{q}}^{(+)}> = (2\pi)^{3} \delta(\mathbf{k}-\mathbf{q})
\end{equation}

The equation that determines the dual state is then easily derived from the Lippmann-Schwinger equation of the 
physical state and the above orthonormality condition

\begin{equation}
<\widetilde{\Psi}_{\mathbf{k}}^{(+)}|\mathbf{q}>
+<\widetilde{\Psi}_{\mathbf{k}}^{(+)}|G_{0}^{(+)}(E_{q})U(E_{q})|\Psi_{\mathbf{q}}^{(+)}> = 
(2\pi)^{3} \delta(\mathbf{k}-\mathbf{q})
\end{equation}

The above equation reduces to the usual one employed in \cite{KA91} in defining the modified DWBA amplitudes in the FKK, 
MSD cross section, if $U$ is taken to be energy independent, viz

\begin{equation}
<\widetilde{\Psi}_{\mathbf{k}}^{(+)}|\mathbf{q}> + <\widetilde{\Psi}_{\mathbf{k}}^{(+)}|U G_{0}^{(-)}(E_{k})|\mathbf{q}> =(2\pi)^{3} \delta(\mathbf{k}-\mathbf{q})
\end{equation}

In general, however, the energy dependence of $U$ has to be taken into account and Eq. (21) has to be dealt with. To exhibit
the physical quantities that appear in the integral equation that determins the dual sate, we use the spectral decompsotion of the free Green function and use the definition of the half-on-energy-shell $T$-matrix, 

\begin{equation}
<\mathbf{k^{\prime}}| T(E_{q})|\mathbf{q}> = <\mathbf{k^{\prime}}|U(E_{q})|\Psi_{\mathbf{q}}^{(+)}>,
\end{equation}

to obtain

\begin{equation}
<\widetilde{\Psi}_{\mathbf{k}}^{(+)}|\mathbf{q}> + 
\int \frac{d\mathbf{k^{\prime}}}{(2\pi)^{3}}<\widetilde{\Psi}_{\mathbf{k}}^{(+)}|\mathbf{k^{\prime}}>\frac{1}{E_{q} - E_{k^{\prime}} +i\epsilon}<\mathbf{k^{\prime}}| T(E_{q})|\mathbf{q}> =
                                                                  (2\pi)^{3} \delta(\mathbf{k}-\mathbf{q})
\end{equation}

It is thus clear that one needs the complete knowledge of the half-on-enegy-shell physical $T$-matrix in order to solve the integral equation to obtain the dual state. Even if it is obtained, one would have to worry about the spectral decomposition of the intermediate channel Green function. For this reason, I would suggest the introduction of an equivalent energy-independent optical potential, $\overline{U}$ which, though highly non-local, can be employed in an integrodifferential equation for both the physical state ( with $\overline{U}$) and the dual state ( with  $\overline{U}^{\dagger}$). The definition of $\overline{U}$ is just

\begin{equation}
U(E_{k})|\Psi_{\mathbf{k}}^{(+)}> = \overline{U}|\Psi_{\mathbf{k}}^{(+)}>
\end{equation}

The obtention of $\overline{U}$ has been discussed during the 80's. I refer to \cite{HMo84} for a dicussion of the contruction and limitation of such an equivalent energy-independent optical potential.

\newpage

\section{Conclusions}

Absorption and fluctuations go hand in hand in nuclear reactions. Initial state fluctuations give rise to the compound nucleus, Hauser-Feshbach, cross section. Intermediate state fluctuations could add contributions to multistep processes.
Final state fluctuations give rise to incomplete fusion. In this talk I have elaborated on these effects. In particular, 
the Surrogate Method, which is incomplete fusion followed by the decay of the xA subsystem, is discussed and means to
assess its limitations have been laid out. The question which theorists should address is why does this method work so
well, as we have heard from several speakers in this Workshop. Further, the multistep direct reaction theory has also
been considered. I have pointed out means to improve the convergence of the FKK version of the theory by including
the intermediate state fluctuations and through a more careful treatment of the intermediate channels' optical potentials and the dual states.


\begin{references}

\bibitem{HM85} M. S. Hussein and K. W. McVoy, Nucl. Phys. A 445, 124 (1985).
\bibitem{Au87} N. Austern et al. Phys. Rep. 154, 125 (1987).
\bibitem{KM79} A. K. Kerman and K. W. McVoy, Ann. Phys. (N.Y.) 22, 197 (1979).
\bibitem{Par95} W. E. Parker et al., Phys. Rev. C 52, 252 (1995).
\bibitem{HFM90} M. S. Hussein, T. Frederico and R. C. Mastroleo, Nucl. Phys. A 511, 269 (1990).
\bibitem{Fr91} T. Frederico, R. C. Mastroleo, B. V. Carlson and M. S. Hussein, J. Phys.G 17, L139 (1991)
\bibitem{Ud84} T. Udagawa, X-H, Li and T. Tamura, Phys. Lett. B 143, 15 (1984).
\bibitem{HMa89} M. S. Hussein and R. C. Mastroleo,  A 491, 468 (1989).
\bibitem{Fr90} T. Frederico, B. V. Carlson, M. C. Mastroleo, L. Tomio and M. S. Hussein, Phys. Rev. C 42, 138 (1990).
\bibitem{Can98} L. F. Canto, R. Donangelo, L. M. de Matos, M. S. Hussein and P. Lotti, Phys. Rev. C 58, 1107 (1998).
\bibitem{TB03} S. Typel and G. Baur, Ann. Phys. (N.Y.) 305, 228 (2003)
\bibitem{Ple05} C. Plettner et al. Phys. Rev. 71, 051602(R) (2005).
\bibitem{ED06} J. Escher and F. S. Dietrich, Phys. Rev. C 74, 054601 (2006).
\bibitem{Esc07} J. Escher, Contribution to this workshop.
\bibitem{FZ77} H. Feshbach and M. Zabek, Ann. Phys. (N.Y.) 107, 110 (1977).
\bibitem{Car99} B. V. Carlson, L. F. Canto, S. Cruz-Barrios, M. S. Hussein and A. F. R. de Toledo Piza, Ann. Phys. (N.Y.)
279, 111 (1999).
\bibitem{CH99} B. V. Carlson, M. S. Hussein, L. F. Canto and A. F. R. de Toledo Piza, Phys. Rev. C 60, 014604 (1999).
\bibitem{Can99} L. F. Canto, B. V. Carlson, M. S. Hussein and A. F. R. de Toledo Piza, Phys. Rev. C 60, 064624 (1999).
\bibitem{Hu02} M. S. Hussein, B. V. Carlson, L. F. Canto and A. F. R. de Toledo Piza, Phys. Rev. C 66, 034615 (2002)
\bibitem{GW01} J. Z. Gu and H. A. Weidenm\"uller, Nucl. Phys. A 690, 382 (2001).
\bibitem{FKK80} H. Feshbach, A. K. Kerman and S. E. Koonin, Ann. Phys. (N.Y.) 125, 429 (1980).
\bibitem{KKM73} M. Kawai, A. K. Kerman and K. W. McVoy, Ann. Phys. (N.Y.) 75, 156 (1973).
\bibitem{VZW85} J. J. M. Verbaarschot, M. R. Zirnbauer and H. A. Weidenm\"uller, Phys. Rep. 129, 387 (1985).
\bibitem{PB62} F.G. Perey and B. Buck, Nucl. Phys. 32, 253 (1962).
\bibitem{HB82} M. S. Hussein and R. Bonetti, Phys. Lett. B 112, 189 (1982).
\bibitem{KA91} A. J. Koning and J. M. Akkermans, Ann. Phys. (N.Y.), 208, 216 (1991).
\bibitem{Ar95} G. Arbanas, M. B. Chadwick, F. S. Dietrich and A. K. Kerman, Phys. Rev. C 51, R1078 (1995).
\bibitem{TUL83} T. Tamura, T. Udagawa and H. Lenske, Phys. Rev. C 26 397 (1983).
\bibitem{NWY88} H. Nishioka, H. A. Weidenm\"uller and S. Yoshida, Ann. Phys. (N.Y.) 183, 166 (1988) 
\bibitem{HMo84} M. S. Hussein and E. J. Moniz, Phys. Rev. C 29, 2054 (1984).

\end{references}
\end{document}